\documentstyle[preprint,aps,prb]{revtex}

\def\ket#1{\vert#1\rangle}
\def\bra#1{\langle#1\vert}
\def\braket#1#2{\langle#1\vert#2\rangle}
\def\br{{\bf r}}
\def\bk{{\bf k}}

\def\pmb#1{\setbox0=\hbox{#1}%
 \hbox{\kern-.025em\copy0\kern-\wd0
 \kern.05em\copy0\kern-\wd0
 \kern-0.025em\raise.0433em\box0} }

\def\taub{{\pmb{$\tau$}}}

\begin{document}
\draft
\title{Complex Band Structures and Decay Length in Polyethylene Chains}
\author{Fabien Picaud,\footnote{Fellow of TMR FULPROP, EU Contract 
ERBFMRXCT970155.}$^1$ Alexander Smogunov,$^{1,2}$ Andrea Dal Corso,$^{1,2}$
 and Erio Tosatti$^{1,2,3}$}
\address{$^{1}$SISSA, Via Beirut 2/4, 34014 Trieste (Italy).}
\address{$^{2}$INFM-DEMOCRITOS, National Simulation Center, I-34014 Trieste
(Italy).}
\address{$^{3}$ICTP, Strada Costiera 11, 34014 Trieste (Italy).}

\date{\today}
\maketitle
\begin{abstract}
The complex band structure of an isolated polyethylene chain is calculated 
within Density Functional Theory (DFT). A plane wave basis and ultrasoft 
pseudopotentials are used. The results are compared with those obtained 
via a local basis set. We obtain a gap between the highest occupied 
molecular orbilar (HOMO) and the antibonding unoccupied molecular 
orbitals of 9.3 eV and a non-resonant tunneling $\beta$ 
parameter of 0.9 per monomer, in reasonable agreement with experiment
and with results obtained via local basis. Polyethylene is a negative 
electron affinity material and the actual gap should be the energy 
of the HOMO with respect to the vacuum level (in DFT approximation 
only about 5.14 eV).  
The Bloch states at imaginary $k$ are mainly free-electron-like 
parabolic bands which are missing in the local basis.
We present also the complex bands of the bulk polyethylene in order
to estimate the effects of the chain-chain interactions on the complex
band structure.
The relevance of these results for the tunnelling conduction of n-alkane 
chains is discussed.


PACS: 71.20-Rv Band structure of polymers and organic compounds. 
      71.15-m Methods of electronic structure calculation.
      73.63-b Electron transport in nanoscale materials and structures.
\end{abstract}

\section{Introduction}

The conductance of interfaces, tunnel junctions or molecular chains connecting 
two tips is of paramount importance for areas such as 
nanoelectronics or molecular electronics.  It is well known that in
order to understand
the electronic and transport properties of surfaces and interfaces, 
the complex electronic band structure (CBS) of the constituent 
materials~\cite{Heine} maybe as important as the real band structure.
However, ab-initio calculation of CBS are still rare.

Very recently the Bloch functions with complex wavevector ${\bf k}$ 
have been revived for the calculation of the ballistic conductance of 
carbon nanotubes,~\cite{Choi} because they provide an effective way
to deal with open quantum systems and transmission 
coefficients within the Landauer theory. 
Furthermore, the spin dependent conductance of 
ferromagnetic/insulator/ferromagnetic junctions has been explained 
in relation with the complex bands of the insulator.~\cite{Mavropoulos}
Inside an isolated surface or tip, the electronic wave functions 
can be written as a superposition of Bloch functions, including
both real and complex ${\bf k}$, namely both propagating and evanescent waves. 
The vacuum tunneling between two tips arises as the overlap in the vacuum 
gap of two opposite evanescent waves.
The electronic states of a nanowire, a nanotube, a long polymer 
bridged between two tips, or of an insulating thin film sandwiched 
between two metallic contacts, can all be expanded in Bloch states
with both real and complex ${\bf k}$.
A detailed analysis of the CBS itself helps to explain
the properties of the tunneling current.
For instance, in the theory of Scanning Tunnelling Microscope (STM), 
the exponential decay of the current with the tip surface distance, 
depends on the metal work function which controls the decay length of
the evanescent metal states in the vacuum.

This concept was applied in Ref.~\onlinecite{Sankey} to calculate
the non-resonant tunneling parameter $\beta$ of n-alkane chains  
measured accurately in recent experiments.~\cite{Wold,Cui} 
The resistance of alkanethiol chains of variable monomer numbers 
was measured by adsorbing the n-alkane chains on a gold (111) 
surface and contacting the other extreme with an Atomic Force 
Microscopy (AFM) tip coated with gold.
The pressure on the chains is kept constant by controlling the force 
exerted by the tip on the n-alkane film.
The resistance depends strongly on the contact force, but at fixed 
force, it depends exponentially on the chain length 
($R=R_0 e^{\beta N}$, where $N$ is the number of CH$_2$ monomers). 
The reported values of $\beta$ range from 0.8 to 1.2 per monomer,
depending on the contact force.
Note that in vacuum, $\beta=\sqrt{{8m \over\hbar^2}\Phi_{Au}}a=3.0$ where
($a=1.25$ \AA\ is the length of a monomer, $\Phi_{Au}=5.3$ eV is the gold 
work function, $m$ is the electron mass, and $\hbar$ the Planck's constant). 
The metal wavefunctions decay exponentially in the polymer and,
at low tip-surface voltage, the current is due to non-resonant 
tunneling because the Fermi energy of the gold contacts lies within 
the gap of the n-alkane chain.~\cite{Sankey} 
The decay length $\beta=2 ({\rm Im} k) a$ 
can be derived from the CBS of a polyethylene
chain (the polymeric form of the n-alkane). 
For sufficiently long chains, only the slowest
decaying state dominates the tunneling current.~\cite{Sankey}
In the CBS, this state belongs to a loop that joins, at imaginary $k$, the 
highest occupied 
molecular orbital (HOMO) and an antibonding unoccupied molecular 
orbital (that we call here LUMO) of the polyethylene chain. Taking the Fermi 
level at the point
where ${d\epsilon\over dk}=0$ a value of $\beta$ close to 1.0 
was found in literature,~\cite{Sankey} based on CBS calculations 
done with two different local basis sets (a minimal one, $s$ for H and
$sp^3$ for C, and an extended one, $sp^3$ for H and $sp^3d^5$ for C). 
There, the decay length $\beta$ was well converged with respect 
to the basis size, but the imaginary bands show a huge basis set 
dependence. This is the aspect that we address here.
A local basis should work very well for the localized bonding and 
antibonding states of molecules, but problems could arise when considering 
saturated, wide gap molecular wires such as n-alkane chains. Here the 
antibonding empty states often fall above the zero of
the electrostatic potential in the vacuum. 
It is known that, experimentally~\cite{Less} as well as 
in DFT,~\cite{Serra,Righi} polyethylene is a negative electron 
affinity material. While holes in the highest valence band are intra-chain
bonding states, electrons in the lowest conduction band are delocalized
inter-chain states while the empty antibonding states lie above the 
vacuum level.
In a monolayer of n-alkane chains attached to the gold surface, 
both intrachain and interchain states appear. The latter cannot 
easily be described and are sometimes missing altogether in local basis 
calculations. 
Ignoring these interchain states or vacuum states in conductante calculations  
may still work or it may not: our point is that they cannot be a priori
neglected. Vacuum states between the HOMO and the LUMO, in general, will
influence the complex band structure of the polymer in a way that needs 
to be addressed.

To clarify this point, we decided to study the complex electronic 
structure of an isolated periodic one-dimensional polyethylene chain
and to compare the CBS obtained with a plane wave basis
with those given by a local basis.\cite{Sankey}
Our results confirm a very strong dependence of CBS upon the choice 
of the basis.
In the plane wave basis, 
the decay length for non-resonant tunneling is shown to be close to that 
obtained by the local basis, 
but, connecting to the free-electron-like states present in the middle 
of the HOMO-LUMO gap, at imaginary $k$, a continuum of free-electron-like 
parabolic states appears. It is very difficult to describe these 
delocalized states with a local basis and in general they are 
substituted by a few basis dependent states.  Depending on the chemical 
potential, they might be relevant or not 
to conductance. For instance, they represent vacuum tunneling 
processes for very short
nanocontacts. Whatever the case, in order to discuss the tunneling current, 
it is useful to keep in mind that they exists and that they might
be important in some cases. 

This paper is organized as follow. In Section II, we describe 
the method used to get the CBS of the polyethylene
with a plane wave basis.~\cite{Smogunov} In Section III, we describe 
the geometry of the isolated chain and of the bulk polyethylene 
used in the calculation and the other
technical details. In Section IV, we present our calculated CBS,
compare with previous calculations, and discuss the consequences 
of our results on the tunneling current of n-alkane chains. 
Finally, Section V contains our conclusions.

\section{Method}

In this Section, we describe briefly the method used to compute the CBS 
of a periodic system. The Bloch theorem allows us to label the one electron 
states with a wavevector ${\bf k}$ and a band index $n$. 
We are interested in Bloch functions with both real and complex 
${\bf k}$. 
Within DFT in the local density approximation (LDA) these Bloch functions
are solutions of the one-electron Kohn-Sham equations. To deal with 
ultrasoft pseudopotentials,~\cite{Vanderbilt} we need to consider the 
generalized form of these equations~\cite{Laasonen} 
(atomic units $e^2=2m=\hbar=1$ are used): 
\begin{equation} 
E\hat S\ket{\psi_{\bf k}}=\left[-\nabla^2+V_{\rm eff}+ 
\hat V_{NL}\right]\ket{\psi_{\bf k}}, 
\end{equation} 
where $V_{\rm eff}$ is the effective local potential 
(see Ref.~\onlinecite{Vanderbilt,Laasonen}), 
$V_{NL}$ is the nonlocal part of the ultrasoft pseudopotential: 
\begin{equation} 
\hat V_{NL}=\sum_{Imn}D^I_{mn}\ket{\beta^I_m} 
\bra{\beta^I_n}, 
\end{equation} 
constructed using the set of projector functions $\beta^I_m$ associated 
with atom $I$ and $\hat S$ is the overlap operator: 
\begin{equation} 
\hat S=1+\sum_{Imn}q^I_{mn}\ket{\beta^I_m} 
\bra{\beta^I_n}.
\end{equation}
The coefficients $q^I_{mn}$ are the integrals of 
the augmentation functions defined in Ref.~\onlinecite{Vanderbilt,Laasonen}. 
It is convenient to rewrite Eq.~(1) in the form: 
\begin{equation} 
E\ket{\psi_{\bf k}}=\left[-\nabla^2+V_{\rm eff}\right]\ket{\psi_{\bf k}}+ 
\sum_{Imn}(D^I_{mn}-Eq^I_{mn})\ket{\beta^I_m}\braket{\beta^I_n} 
{\psi_{\bf k}}. 
\label{sequation} 
\end{equation} 
In this equation we take the energy $E$ as a fixed external parameter 
and the values of ${\bf k}$ (in general complex numbers) are determined 
so that $\ket{\psi_{\bf k}}$ obeys the Bloch periodicity condition. 
This problem, formulated within the ultrasoft pseudopotential scheme 
is not substantially different from the norm conserving one solved 
in Ref.~\onlinecite{Choi}.

We solve this equation in a supercell geometry along the $x$ and $y$ 
directions.  A wave propagating at energy $E$ will have $\bk=(\bk_\perp,k)$.
Due to the supercell geometry $\psi_{({\bf k}_\perp,k)}$ will have the 
Bloch form along $x$ and $y$:
\begin{equation}
\psi_{({\bf k}_\perp,k)}(\br_{\perp}+{\bf R}_\perp,z)=e^{i{\bf k}_\perp
{\bf R}_\perp} \psi_{({\bf k}_\perp,k)}(\br_\perp,z),
\label{uno}
\end{equation}
where ${\bf k}_\perp$ is a real vector whereas $k$ can be complex.
Different ${\bf k}_\perp$ do not mix and can be considered separately.
From now on, we omit the ${{\bf k}_\perp}$ on 
$\psi_{({\bf k}_\perp,k)}$. 

As in Ref.~\onlinecite{Choi} we write the solution of the integro-differential 
Eq.~(\ref{sequation}) as  a linear combination:   
\begin{equation} 
\psi_k(\br)=\sum_n a_{n,k}\psi_n(\br)+\sum_{Im} c_{Im,k} 
\psi_{Im}(\br), 
\label{tre} 
\end{equation} 
where $\psi_n$ are linearly independent solutions of the homogeneous equation: 
\begin{equation} 
E\psi_n(\br)=\left[-\nabla^2+V_{\rm eff}(\br)\right]\psi_n(\br), 
\end{equation} 
and $\psi_{Im}$ is a particular solution of the inhomogeneous equation: 
\begin{equation} 
E\psi_{Im}(\br)=\left[-\nabla^2+ 
V_{\rm eff}(\br)\right]\psi_{Im}(\br)+\sum_{{\bf R}_\perp} 
e^{i{\bf k}_\perp {\bf R}_\perp} 
\beta^I_m(\br-\taub^I-{\bf R}_\perp).
\end{equation} 
Both $\psi_n$ and $\psi_{Im}$ are $(x,y)$ periodic as in
Eq.~(\ref{uno}).  Summation over $Im$ in Eq.~\ref{tre} involves 
all the projectors in
the unit cell  and the coefficients $c_{Im,k}$ are determined by: 
 \begin{equation} 
 c_{Im,k}=\sum_n [D^I_{mn}-Eq^I_{mn}] 
 \int [\beta^I_n(\br-\taub^I)]^*\psi_k(\br)d^3r. 
\label{cin}
 \end{equation} 
Note that the set 
of wave functions $\psi_n$ is infinite. In practice the 
expansion $\psi_n(\br)=\sum_{{\bf G}_\perp}\psi_n({\bf G}_\perp,z) 
e^{i({\bf k}_\perp+{\bf G}_\perp)\cdot\br_\perp}$ over two-dimensional 
plane waves is used. If one considers only $N_{2D}$ plane waves with 
${\bf G}_\perp^2\leq E_{\rm cut}$ the number of $\psi_n$ becomes finite 
and equals to $2N_{2D}$. As in Ref.\onlinecite{Choi}, to find the 
functions $\psi_n$ 
we discretize the unit cell along the $z$ axis by dividing it 
into slices and replacing the $V_{\rm eff}(\br)$ in each slice 
by a $z$-independent potential. The $\psi_n$ in the slices is a 
linear combination of two exponentials with coefficients obtained 
by a matching procedure.

The allowed values of $k$ for a given energy $E$ can be determined by 
imposing periodicity along $z$ to the Bloch function and to its 
$z$ derivative. If $d$ is the cell length along $z$ we have:
\begin{equation} 
 \psi_k({\bf G}_\perp,d)=e^{i k d} 
 \psi_k({\bf G}_\perp,0).
\label{dod}
\end{equation} 
\begin{equation} 
\psi'_k({\bf G}_\perp,d)=e^{i k d} 
\psi'_k({\bf G}_\perp,0).
\label{tredi}
\end{equation} 
Inserting Eq.~\ref{tre} into Eqs.~\ref{cin}-\ref{tredi} one can show
that the last three equations are equivalent to the generalized eigenvalue 
problem: 
\begin{equation}
 AX=e^{i 
 k 
 d}BX,
\end{equation} 
where $A$ and $B$ are known matrices.
We solve this problem to obtain in general a complex $k$ and coefficients 
$X=\Big\{a_{n,k},c_{Im,k}\Big\}$ for the generalized Bloch state 
$\psi_k$ at a given energy $E$ and $\bk_\perp$. 

Our CBS calculations proceed in two steps. First we compute 
with a standard electronic structure code ({\tt PWSCF}) \cite{PWSCF} 
the ground state electronic structure of the system and we obtain 
the effective local potential $V_{\rm eff}(\br)$ and the screened
coefficients $D^I_{mn}$. 
In a second step, we use the potential $V_{\rm eff}(\br)$ and the screened
coefficients $D^I_{mn}$ to compute the CBS applying 
the scheme described above.

\section{Geometrical model and technical details}

The n-alkane chains [X(CH$_{2}$)$_{n}$Y] are made of CH$_2$ monomers
with singly bonded carbon atoms and two terminating groups (indicated with
X and Y). If grafted via a thiolic terminating group (X=SH) on 
the Au(111) surface, 
they can spontaneously assemble into a dense and ordered monolayer 
film to form an insulating layer.~\cite{Poirier}
We model the n-alkane chain with a periodic polyethylene chain
[poly-CH$_{2}$] as depicted in Fig.\ref{fig1}a. 
Since we shall compare our results with Ref.~\onlinecite{Sankey}, we choose
the same geometry, C-C-C and H-C-H angles are those of the 
perfect tetrahedral structure (109.5$^\circ$) and the C-C bond 
lengths and the C-H bond lengths 
are respectively equal to 1.54 \AA\ and 1.10 \AA. 

The chain is modelled with a tetragonal supercell. The chain is parallel 
to the $z$ axis and periodically repeated along the $x$ and $y$ directions 
with a chain-chain distance $a$ ($10.58$ \AA) large enough to avoid 
replica interactions (tests with a larger supercell gave 
no relevant change of the results).  
The length of the unit cell $c$ in the $z$ direction is $c=2.51$ \AA 
(experimental $c=2.50$ \AA).
In order to estimate the influence of the 
chain-chain interaction, we also calculated bulk polyethylene. 
The rombohedral unit cell edges are $a=4.93$ \AA, $b=7.40$ \AA, 
$c=2.534$ \AA, as in Ref.~\onlinecite{Serra}, and the internal 
coordinates were allowed to relax. 
We found $d_{CC}=1.51$ \AA, $d_{CH}=1.11$ \AA, 
$\alpha_{CCC}=114^\circ$ and $\alpha_{HCH}=105^\circ$ (See Fig.~\ref{fig1}b)
to be compared with experimental values $d_{CC}=1.53$ \AA, 
$d_{CH}=1.09$ \AA, $\alpha_{CCC}=113.3^\circ$ and $\alpha_{HCH}=109.5^\circ$.

We calculated the electronic structure of these systems with the {\tt PWSCF} 
code~\cite{PWSCF} using the local density approximation (LDA) with the 
parameterization of Perdew and Zunger.~\cite{Perdew}  The electron-ion 
interaction was described by ultrasoft pseudopotentials.~\cite{PWSCF} 
The wavefunctions were expanded in a plane waves basis with an energy cut-off
of $25$ Ry. The cut-off for the charge density is $150$ Ry.
The integrations on the Brillouin zone 
are done with a uniform mesh of $40$ $k$ points for the isolated chain
and with a mesh $3\times3\times10$ ${\bf k}$ points for the bulk.

The CBS calculation was subsequently carried out with the technique 
described above. The unit
cell length $c$, in the $z$ direction, is divided into 15 slices and 
the integrals are performed with 11 points in each slice (the slice
thickness is 0.167 \AA\ which is sufficient to describe accurately 
the slowly varying effective potential).
The two dimensional cut-off energy is taken as in the electronic structure
calculation. We consider only the ${\bf k}_\perp=\Gamma$ point. This is a 
very good approximation for the isolated chain, while it gives 
the complex band structure of bulk polyethylene only along one line in
reciprocal space.

\section{Results and discussion}
The CBS calculated for the periodic isolated polyethylene chain and for the 
bulk polyethylene are shown in Fig.~\ref{fig2}a 
and \ref{fig2}b respectively. The lowest unoccupied state, identifiable with
the vacuum zero in the limit of infinite interchain spacing, is 
taken as the zero of the energy.
In each Figure, the right panel shows the bands at the real $k$, 
the Bloch modes propagating along the $z$ direction,
whereas the left panel displays the bands obtained for complex $k$, 
Bloch modes evanescent (or exponentially diverging) in the $z$ direction 
(note that for each solution $k$, at a given energy $E$, also the complex 
conjugate $k^*$ is a solution). Bands with purely imaginary $k$ 
are represented with small dots, while bands with complex $k$ are 
projected on the imaginary $k$ plane and indicated by small circles.

The energy bands at real $k$ of polyethylene (both for the bulk structure
and for an isolated chain) have been reported in several papers.
\cite{Sankey,Serra,Miao,Montanari,Martins,Seki,Ueno,Springborg,Sule,Sun} 
The valence bands, C-C and C-H bonding states, have been also compared 
with experimental angular-resolved photoemission 
data.~\cite{Miao,Seki,Springborg,Sun} 
In Table~I, we compare our results with previous DFT calculations.
All theoretical band widths are in substantial agreement with each other,
weakly dependent on the basis set used or the actual geometry (isolated chain 
or bulk polyethylene).
Our conduction band widths are 14.0 eV (chain) and 13.6 eV (bulk) 
in good agreement with previous calculations.
All these results are also in reasonable agreement with the experimental 
value of 16.2 eV.\cite{Seki,Ueno,Fujimoto}
In all theoretical calculations, however, the absolute 
position of the occupated bands of the chain with respect to the vacuum level, 
is too high if compared with experiment. In the bulk, according to the  
photoemission data of Ref.~\onlinecite{Fujimoto}, the HOMO level lies 10 eV 
below vacuum (8.8 eV below vacuum, according to Ref.~\onlinecite{Less}). 
DFT calculations however place the valence band top
about 6.0 eV below vacuum. Thus, for example, a shift of 4.5 eV was
needed in Ref.~\onlinecite{Miao} to bring theory and experiment in 
closer agreement.

The theoretical band gaps appear to be in disagreement with each other.
For the chain there are two sets of values: depending on the 
exact geometry or exchange and correlation functional,  
gaps calculated with plane waves cluster around 5-6 
eV,~\cite{Serra,Montanari,Martins} 
while gaps calculated with local basis are around 8-10 eV.~\cite{Sankey,Miao} 
We find an energy gap of 5.14 eV (for the isolated chain) 
which becomes 6.4 eV (in bulk polyethylene) in agreement with previous 
plane waves results.~\cite{Serra,Montanari,Martins}
S\"ule et al.~\cite{Sule} have found that the band gap depends 
on the basis-set quality. Variations of several electron volts 
have been attributed to the appearance in the gap of new states 
belonging to the continuum when the 
basis-set size is increased. 
The nature of the conduction band states obtained with plane waves 
has been discussed in detail by Serra et al..\cite{Serra}
In bulk polyethylene, the charge 
density associated to the $\Gamma$ point wavefunction at the
conduction band minimum shows clear maxima between the chains. 
Increasing the interchain distance, this state evolves to a 
free-electron-like state, or perhaps a surface state.~\cite{Righi}
In addition a slab calculation has shown that, within GGA, 
polyethylene is a negative electron affinity material.~\cite{Righi} 
According to experiment (Ref.~\onlinecite{Less}) the electron conduction 
band lies at an energy $A$ above the vacuum level, and $A$ is about 
0.5 eV above the vacuum.
It is a characteristic feature of DFT in the LDA approximation to yield
gaps that are generally too small, with the exception of bonding 
anti-bonding gaps.
Comparing our complex bands with those obtained via a localized 
basis,~\cite{Sankey} it is possible to identify the reason of the different 
theoretical values for the band gap.
In the chain, the gap calculated by plane waves measures the position 
of the HOMO with respect to the vacuum level, while the gap 
calculated by local basis is very close to the energy of the HOMO with 
respect to an antibonding empty state (that we call LUMO).
Actually free-electron-like states are poorly described by a local 
basis set and they barely appear within the HOMO-LUMO gap. 
Instead in our calculation, the lowest conduction bands are parabolic, 
indicating free-electron-like states which, in bulk polyethylene, indicate 
interchain states.~\cite{Serra,Righi}
In a truly isolated polyethylene chain the vacuum zero energy
identifies the onset of free-electron-like states.
The LUMO, which is higher in energy than the 
vacuum zero should appear as a resonant level in the density of states at
positive energy. Since we are using a supercell geometry the density 
of free-electron-like states will however depend on the box size.

We now analyze the band structure at imaginary $k$ and at complex $k$ of the
isolated chain and compare our results with those of Ref.~\onlinecite{Sankey}. 
At imaginary $k$ the CBS is formed mainly by free-electron-like parabolas. 
The presence of the polyethylene chain splits the degeneracy of 
free-electron bands. In our calculation these bands are discrete, but
at infinite box size, above the free-electron parabola departing from
the $\Gamma$ point at vacuum zero energy, they are a continuum of imaginary
$k$ states.
The continuum of parabolic bands is not well reproduced with a local
basis and actually the states that one obtains at imaginary $k$ depends 
on the basis as shown in Ref.~\onlinecite{Sankey} where a minimal and an 
extended basis give totally different results.
At the $\Gamma$ point parabolic bands depart also from all the valence 
states except the HOMO. 
A loop starts from the HOMO and joins the LUMO of the molecule which is
at $9.3$ eV above the HOMO, 4.13 eV higher, in DFT, than the vacuum.
We note that our HOMO-LUMO gap is correctly close to the local 
basis gap.\cite{Sankey}

For complex $k$ there are loops joining maxima and minima
of the bands. The loop between $-12.3$ eV and $-9.9$ eV closes the gap 
between the C-C $\sigma$-bonding band and the C-H bonding states. Since
these states are well reproduced by local basis, the shape and
the size of this loop is in good agreement with Ref.~\onlinecite{Sankey}.
Two additional loops, also in good agreement with the local basis 
calculation, join maxima in the valence bands with minima in 
the conductions states. 
In the energy gap, these states have a decay length shorter than 
the loop joining the HOMO and the LUMO of the molecule.

The imaginary and complex $k$ bands of bulk polyethylene are shown in 
Fig.~\ref{fig1}b. From the comparison of this result with the isolated
chain, we can estimate the effect of the reduced interchain distance 
on the complex bands. 
As expected, the density of free-electron-like states is much reduced.
Furthermore, since there are four monomers per unit cell, all the bands are 
doubled with small splittings in the valence bands and much larger 
splittings in the conduction bands.
The loop which, in the isolated chain, joins the HOMO and
the LUMO at imaginary $k$ is now split in two loops through an  
avoided crossing with a parabolic band. 
The two loops that, at complex $k$, join valence and conduction
bands are both split and, since the gap is now indirect, in a small
energy window below the LUMO, the states in one of these loops have
a longer decay lenght than those in the loop which joins the
valence and conduction band at the $\Gamma$ point.

In a typical experiment where the tunneling decay length $\beta$ of
an n-alkane chain is measured, the molecule is adsorbed on an Au(111) 
substrate by terminating it with a thiol group. As shown in 
Ref.~\onlinecite{Sankey} the Fermi level of the gold 
substrate is between the HOMO-LUMO gap. Its real position is not
known experimentally; theoretically it depends
on the basis set used in the calculation. With
plane waves, it is at about 3.5 eV~\cite{Sankey} above the HOMO of the polymer. 
It is reasonable to suppose that the presence of the thiol group, as
well as the termination of the chain are perturbations which decay 
much faster than the metal tunneling states. With this hypothesis, the 
typical decay length of the tunneling current can be approximated
with the longest decay length of the evanescent Bloch states 
of the polyethylene chain at the Fermi energy.
The important complex band determining the tunneling decay 
length is therefore the loop joining the HOMO and LUMO states
since it has the smallest ${\rm Im} k$ at the Fermi level. 
Our value for ${\rm Im} k$ is 0.36 \AA$^{-1}$, 
corresponding to $\beta=0.9$ per monomer, in good agreement with 
experiment and with the calculation of Refs.~\onlinecite{Sankey,Wold,Cui}, 
the small difference with Ref.~\onlinecite{Sankey} being due only 
to the choice of the Fermi level.
Our CBS shows that local basis and plane waves give the
same decay lenght for the non-resonant tunneling regime if the
chemical potential is much lower than the vacuum.
According to a local basis calculation, increasing the substrate tip 
voltage, one would however expect to move the chemical potential of the metal 
and to enter in a resonant tunneling regime where the
current pass through the LUMO of the molecule.~\cite{Datta}
According to our plane wave LDA calculation there is instead a range of
voltages before the resonant tunneling regime, where
vacuum tunneling should start to contribute to the current.
The LDA underestimates the voltage necessary for this phenomenum to occur, 
since the HOMO is only 5.14 eV below the vacuum,
while the correct gap should be of 8.8 eV\cite{Less} but,
in negative electron affinity molecules, vacuum states do start to 
play a role before the LUMO.
The LDA error in positioning the HOMO of the 
molecule must be kept in mind before comparing with experiment the
conductance calculated with a plane wave basis.
Nonetheless, negative electron affinity molecules such as the n-alkanes must 
be treated carefully expecially with local bases because the incorrect
description of vacuum states may lead, not only to a complete lack of
free-electron-like states in the HOMO-LUMO gap, but also to the presence of
many fictitious states at imaginary $k$, that are totally dependent on the 
choice of the basis set.

\section{Conclusion}

In this paper, the CBS of a periodic polyethylene chain has been calculated 
using a plane wave basis set and compared with the CBS obtained 
with a local basis.~\cite{Sankey}
We have shown that the tunneling decay length $\beta$ of n-alkane
chains, here replaced by the polyethylene, is independent
of the choice of the basis, at least as long as at low voltages
the tunneling is dominated by a slowly decaying state which belongs to the 
loop joining, at imaginary $k$, the HOMO and LUMO states.
Instead, our results show that the choice of the basis could be critical
in large gap polymers with negative electron affinity particularly at large
voltages. Future calculations of the conductance must be done carefully
because of the presence of free-electron-like vacuum states within the 
HOMO-LUMO gap, signifying vacuum tunneling besides intra-chain tunneling.
With a plane wave basis, the free-electron-like 
modes are taken into account but the density functional approximation
puts the HOMO too close to the vacuum.
With a local basis that treats correctly the bonding and antibonding 
states of the polymer the density functional problem remains, and on
top of that the free-electron-like states are replaced by basis dependent 
levels.

\acknowledgements
{\vspace{-0.4cm} This project is sponsored by MIUR COFIN2001,
INFM/F, INFM/G, Iniziativa Trasversale Calcolo Parallelo, and by
EU through TMR FULPROP, Contract ERBFMRXCT970155.} Part of the
calculations have been done on the IBM SP3  at Cineca in Bologna.

\vfill 
\eject

\begin{figure}
\caption{a) Geometric structure of the polyethylene chain considered 
in this work.  The geometry is kept close to the ideal tetrahedral one, 
with CC and CH bond lengths as in Ref.~\cite{Sankey}.  
b) Geometric structure of bulk polyethylene.}
\label{fig1}
\end{figure}

\begin{figure}
\caption {a) Complex band structure of polyethylene. The right panel
shows the real bands, while imaginary and complex bands are shown in the
left panel. Small dots represent bands with purely imaginary $k$, while
small circles represent Bloch states with complex $k$ projected on the
imaginary plane. b) As in a) for bulk polyethylene.}
\label{fig2}
\end{figure}

\begin{table}
\caption {Band properties of an isolated polyethylene chain and of bulk
 polyethylene calculated in the present work compared with selected
 previous results.}
\begin{tabular}{lccccc}
 &Basis & Functional & Bandwidth (eV)
&  C-C Bandwidth (eV) &Gap (eV)\\
\hline
Chain     &  &  &  &  & \\
This work     & Plane Waves & LDA & 14.0 & 6.2 & 5.14 \\
Ref.~\onlinecite{Serra}     & Plane Waves & GGA & 14.0 & 7.0 & 5.1 \\
Ref.~\onlinecite{Montanari}     & Plane Waves & LDA & 13.5 & 6.8 & 5.7 \\
Ref.~\onlinecite{Montanari}     & Plane Waves & GGA & 13.5 & 6.7 & 6.0 \\
Ref.~\onlinecite{Sankey}     & Local basis & ? & 13.5 & 7.0 & 9.0-10.25 \\
Ref.~\onlinecite{Miao}     & Gaussian basis  & LDA & 14.0 & 6.1 & 8.0 \\
Bulk     &  &  &  &  & \\
This work     & Plane Waves & LDA & 13.6 & 7.0 & 6.4 \\
Ref.~\onlinecite{Martins}     & Plane Waves & LDA & 14.2 & 8.1 & 6.7 \\
Ref.~\onlinecite{Martins}     & Plane Waves & GGA & 13.7 & 7.1 & 6.2 \\
Ref.~\onlinecite{Serra}     & Plane Waves & GGA & 14.0 & 7.0 & 6.4 \\
Exptal     &  &  & 16.2$^a$  & 7.2$^a$ & 8.8$^b$\\
\end{tabular}
$^a$Ref.~\onlinecite{Fujimoto} \\
$^b$Ref.~\onlinecite{Less} \\
\vskip0.2cm
\end{table}

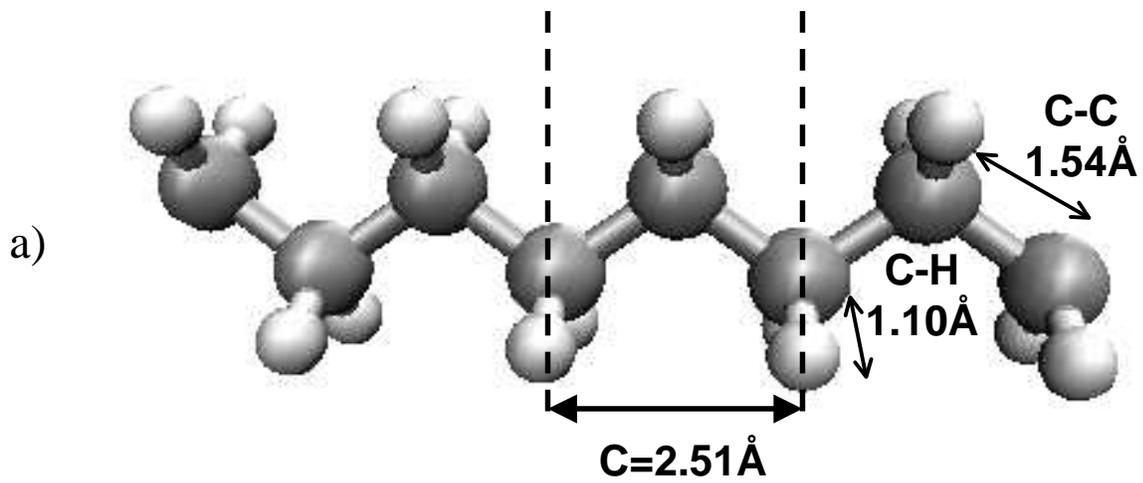
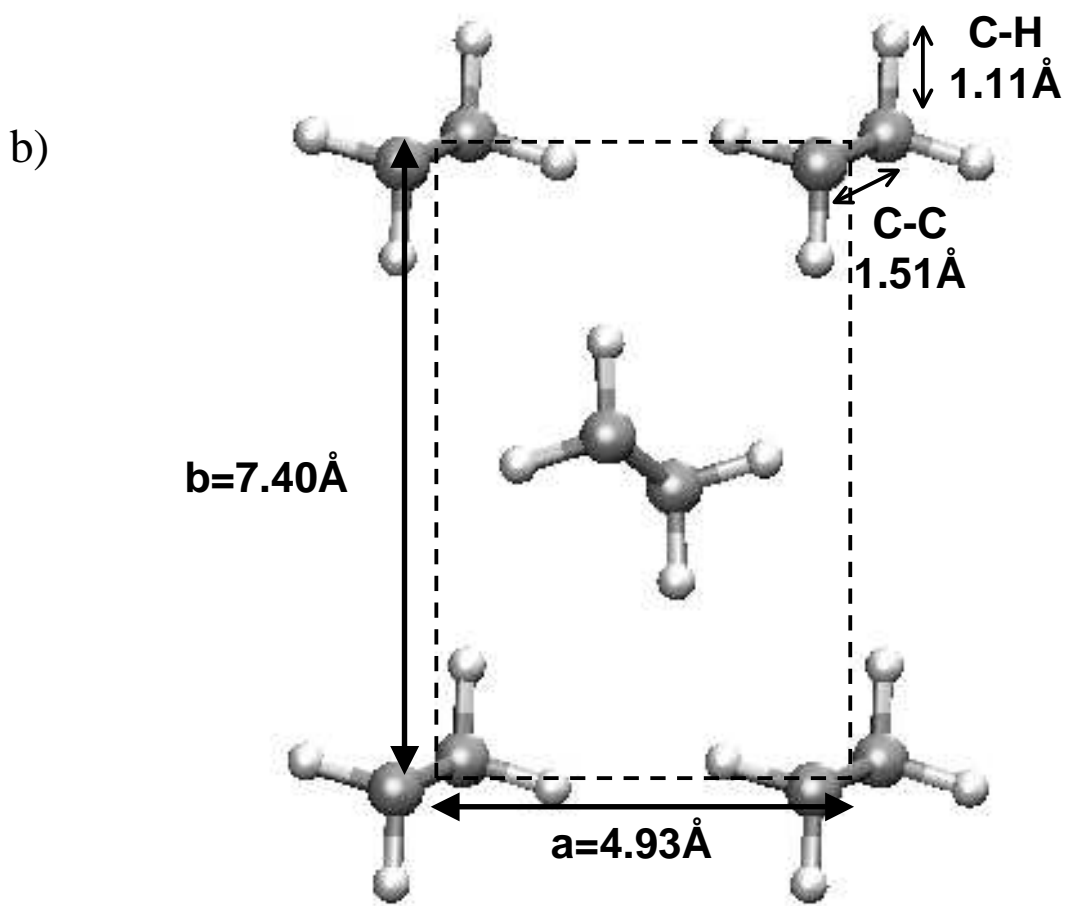
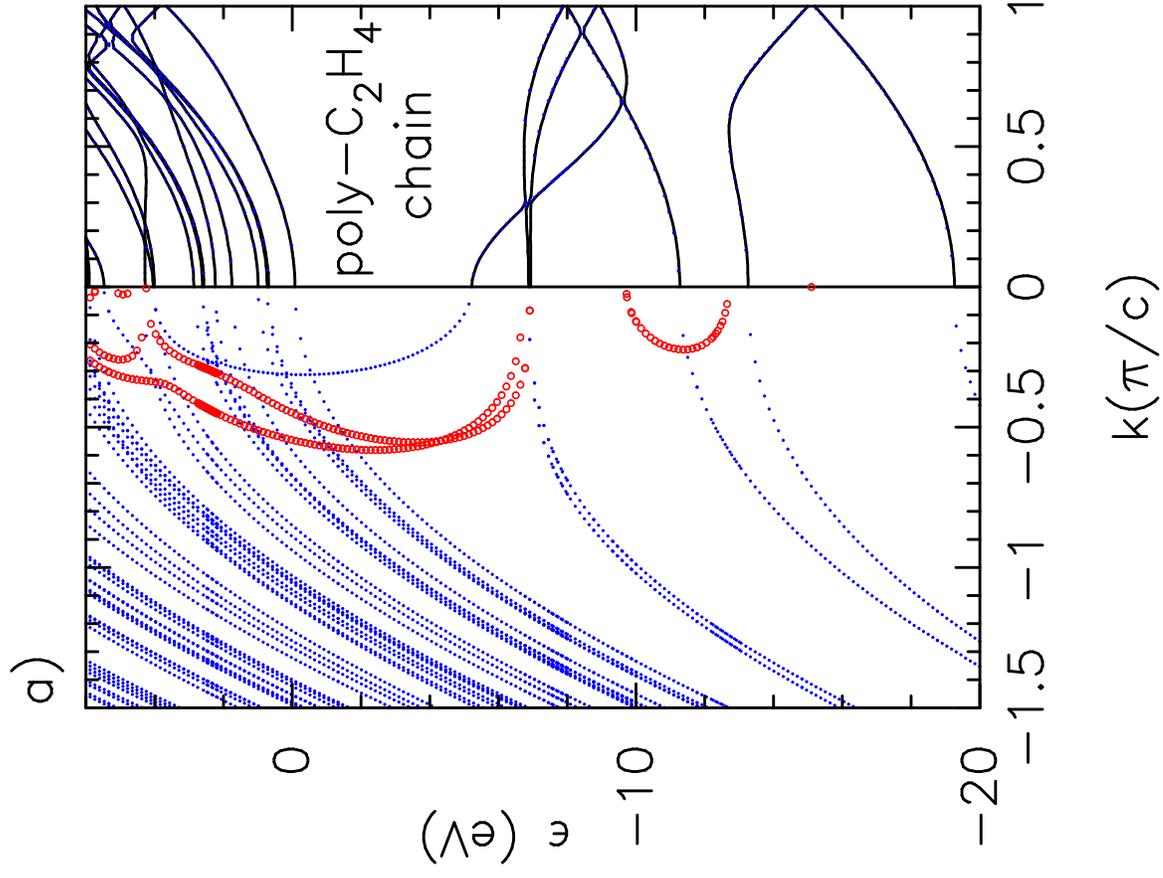
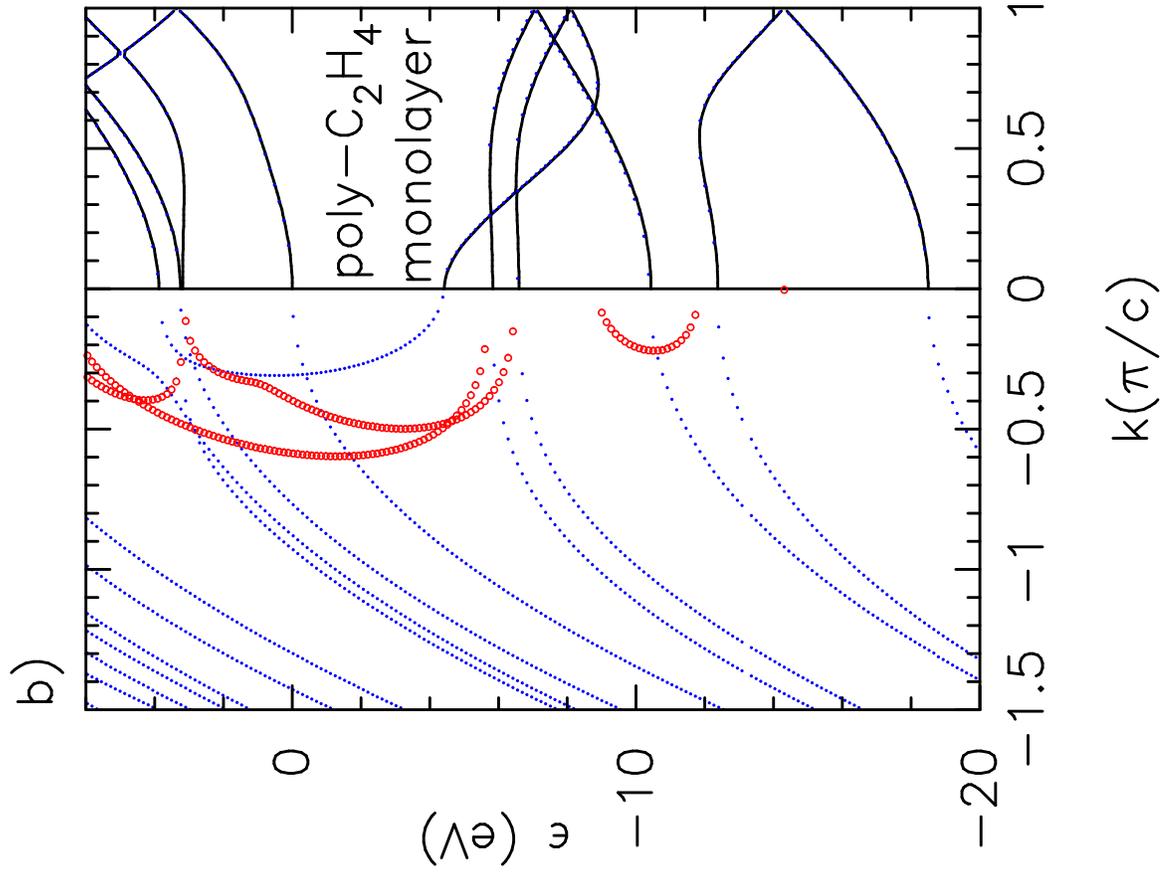
\end{document}